\input epsf
\documentclass[nohyper,notoc]{article} 
\usepackage{multirow}
\usepackage{epsfig}
\usepackage{bm}

\usepackage[utf8]{inputenc} 
\usepackage[T1]{fontenc}    

\def\ben{\begin{enumerate}}
\def\een{\end{enumerate}}
\def\bit{\begin{itemize}}
\def\eit{\end{itemize}}
\def\beq{\begin{equation}}
\def\eeq{\end{equation}}
\def\bea{\begin{eqnarray}}
\def\eea{\end{eqnarray}}
\def\bq{\begin{quote}}
\def\eq{\end{quote}}
\def \lsim{\mathrel{\vcenter
     {\hbox{$<$}\nointerlineskip\hbox{$\sim$}}}}
\def \gsim{\mathrel{\vcenter
     {\hbox{$>$}\nointerlineskip\hbox{$\sim$}}}}
\def\gappeq{\mathrel{\rlap {\raise.5ex\hbox{$>$}}
{\lower.5ex\hbox{$\sim$}}}}
\def\lappeq{\mathrel{\rlap{\raise.5ex\hbox{$<$}}
{\lower.5ex\hbox{$\sim$}}}}

\def\hats{\hat{s}}
\def\hatt{\hat{t}}
\def\hatu{\hat{u}}
\def\epem{e^+e^-}
\def\sigmahat{\hat{\sigma}}

\def\g{\gamma}

\begin{document}

\renewcommand{\thefootnote}{\fnsymbol{footnote}}
\begin{center}
{\Large {\bf
Estimated constraints on  $t$-channel Leptoquark
exchange from LHC contact interaction searches 
}}

\vskip 25pt
{\bf
Assia Bessaa and 
   Sacha Davidson \footnote{E-mail address:
s.davidson@ipnl.in2p3.fr}    } 
 
\vskip 10pt  
{\it IPNL, CNRS/IN2P3,  4 rue E. Fermi, 69622 Villeurbanne cedex, France; 
Universit\'e Lyon 1, Villeurbanne;
 Universit\'e de Lyon, F-69622, Lyon, France
}\\
\vskip 20pt
{\bf Abstract}
\end{center}

\begin{quotation}
  {\noindent\small 
The $t$-channel exchange of a first generation leptoquark 
could contribute to the
cross-section for $q\bar{q} \to \epem$. The leptoquark
is off-shell, so this process can be sensitive to leptoquarks
beyond the mass reach of  pair production searches
at the LHC (currently $m_{LQ}> 830$ GeV). 
We attempt to analytically translate ATLAS bounds on $ (\bar{q}  \g ^\mu q)
(\bar{e}  \g _\mu e) $ contact interactions to the various
scalar leptoquarks, and obtain a bound on their quark-lepton
coupling of order $\lambda^2 \lsim (m_{LQ}/2$ TeV)$^2$.
The greatest difficulty in this translation is that
the leptoquarks do not induce the contact interaction
studied by ATLAS, so the interference with the Standard
Model is different. If bounds were quoted on the
functional dependance of the cross-section on $\hats$, 
rather than  on particular
contact interaction models, this difficulty in applying
experimental bounds to theoretical models could be
circumvented.
\vskip 10pt
\noindent
}

\end{quotation}

\vskip 20pt  

\setcounter{footnote}{0}
\renewcommand{\thefootnote}{\arabic{footnote}}


\section{Introduction}
\label{intro}

The LHC  has  sensitivity to new particles from beyond
its kinematic reach, which could  materialise
as an excess or deficit of  events  at high energy.  Such  
modifications of the high energy tail of distributions
are commonly parametrised by  four-fermion  ``contact interactions'',
with coefficient  $ \pm \frac{4\pi}{\Lambda^2} $. 
Experimental  results are quoted as lower bounds on $\Lambda$,
for a selection of contact interactions.
The question that interests us, is  whether such bounds provide 
 useful constraints on the New Physics which could affect
the tails of distributions.

Concretely, we will consider the partonic process
$q\overline{q} \to e^+e^-$.  A first generation leptoquark
(for reviews, see \cite{LQrev}) 
exchanged in the $t$-channel (see figure
\ref{fig:Feynman})  could mediate this process. 
This  process
 could be sensitive to heavier  leptoquarks  than 
could be pair-produced via  strong interactions at the
LHC (the current bound on  pair-produced first generation leptoquarks
is $m_{LQ} \gsim 830$ GeV\cite{CMS}). However,
this process occurs via the leptoquark-quark-lepton coupling,
so will only be observable for ${\cal O}(1)$ couplings. 
 From the  ATLAS
 bound   \cite{ATLAS} 
on $(\overline{q} \g^\mu P_L q) ( \overline{e} \g_\mu P_L e) $
contact interactions, we  attempt analytically  to estimate bounds on 
the mass and  quark-lepton coupling of the leptoquark.
Two issues will arise. 
First,  in section \ref{sec:CI},  we  treat
the leptoquark exchange as a contact interaction.
However,  none of the seven
possible  leptoquarks
interfere with the SM in the same way as the ATLAS operator.
We  attempt to circumvent 
this problem by assuming the bound comes from the interference term,
and making simple approximations to the Parton Distribution
Functions (pdfs).  
The second hurdle is the leptoquark propagator
$\sim 1/(m_{LQ}^2 + \hatt)$,  which is taken into account
in section \ref{sec:prop}.
As expected, for $m_{LQ}  \lsim 2\sqrt{\hats}$,
the propagator reduces the cross-section, and
therefore weakens the bounds. This effect
seems less significant and
 easier to  estimate that the consequences
of interference.
Section \ref{sec:disc} concludes with  a summary of
the bounds we obtain  on first generation leptoquarks, 
and  a discussion of  the difficulties of
translating contact interaction bounds to any
realistic model without doing a full analysis. 
Were it possible for experimentalists to set bounds on
the functional form of the cross-section, 
it could be easier for  theorists to translate such
limits to their
favourite models.

Combined constraints on leptoquarks, from pair production
and single leptoquark exchange in $s$ and $t$ channels,
has been studied  at HERA (see {\it e.g.} \cite{Pirumov}). 
 Constraints  from LHC contact interaction searches
on  hypothetical new
particles exchanged in the $t$-channel  have previously
been calculated for  $Z'$s \cite{GM}.  Single leptoquark
{\it production} via  $t$-channel diagrams has
also recently been studied in \cite{DFG}.


\section{ The ATLAS analysis, Leptoquarks and Kinematics }
\label{sec:notn}

This section provides a brief overview of
the experimental analysis  we use\cite{ATLAS},
 leptoquarks,   and our notation.

ATLAS searched \cite{ATLAS} for contact interactions of the form
\beq
{\cal L}_{ATLAS} = \eta \frac{4 \pi}{\Lambda^2}
{\Big [}
(\overline{d} \gamma^\mu P_L d ) (\overline{e} \gamma_\mu P_L e  )
+
(\overline{u} \gamma^\mu P_L u ) (\overline{e} \gamma_\mu P_L e  )
{\Big ]}
\label{inpy6}
\eeq
where $\eta= +/- 1$ corresponds to destructive/constructive interference with
$Z/\g$ exchange. The 95 \%  confidance level $(CL)$ bounds 
obtained with  5 fb$^{-1}$ of data at $\sqrt{s} =7$ TeV, are 
\beq
\Lambda_{con} \gsim 11.75 ~{\rm TeV}~~,~~
\Lambda_{des} \gsim 9.3 ~{\rm TeV}~~~.
\eeq

The analysis presents the  number  of
events expected and observed 
as a function of the invariant mass-squared of the $e^+e^-$
pair
\beq
M^2_{\epem} = \hats = x_1 x_2 P_+ \cdot P_- = x_1 x_2  s
\label{M2ee}
\eeq
 in bins of width given in the left colomn of
table \ref{tab:atlas}. 
The background  calculated by  ATLAS
includes   $q\overline{q} \to Z/\gamma \to e^+e^-$,
 as well as other
SM processes such as   $t\overline{t}$ production 
 and dibosons.

\begin{table}[htp!]
\begin{center}
\begin{tabular}{||c||c|c|c||}
\hline   
bin (GeV) & 
 $Z/\g$+ other SM  &all SM +$\Lambda_{cons} = 12$ TeV
&data \\
\hline
400-550 &203 +73 $\pm$ 25 & 293 $\pm$27 &270  \\
550-800 &62 +20 $\pm$ 9 & 96 $\pm$9 &88   \\
800-1200 &12.1+ 3.2  $\pm$ 2 & 23.6 $\pm$2.3 &17 \\
1200-1800 &1.38 +.36  $\pm$ .33 & 5.1 $\pm$.5 &3 \\
1800+ &.085 +  .035$\pm$.04 & 0.87 $\pm$.14 &0  \\
\hline
\end{tabular}
\caption{
In the first colomn,
bins  of dilepton mass $M_{\epem}$
(where $M^2_{\epem} = (p_e+ p_{\overline{e}})^2$).
The following  colomns  are from 
the ATLAS paper \cite{ATLAS}:
 the expected number of events due to the SM, 
due to the SM  plus a contact interaction with $\Lambda= 12$ TeV
and constructive interference with the SM,
and finally the data.
\label{tab:atlas}}
\end{center}
\end{table}

We consider    scalar leptoquarks, 
with renormalisable $B$ and $L$
conserving interactions, which generate
an interaction between first generation
quarks and electrons. 
In the notation of
Buchmuller,Ruckl and Wyler\cite{BRW}, the
leptoquark interactions with
quarks and leptons 
can be added to the SM Lagrangian  as:
\bea
{\cal L}_{LQ} & = & 
S_0 ( {\lambda}_{L S_0} \overline{\ell} i \tau_2 q^c
+ { \lambda}_{R S_0} \overline{e} u^c ) +
\tilde{S}_0 { \tilde{\lambda}}_{R \tilde{S}_0} 
\overline{e} d^c   \nonumber \\ &&
+
S_{2} ( { \lambda}_{L S_2} \overline{\ell}  u
+ { \lambda}_{R S_2} 
\overline{e } q [i \tau_2  ]) +
\tilde{S}_2 { \tilde{\lambda}}_{L \tilde{S}_2} 
\overline{\ell} d   \nonumber \\ &&
 + \vec{S}_1 { \lambda}_{LS_{1}} \overline{\ell} i\tau_2 \vec{\tau}
q^c \cdot  + h.c.
\label{BRW}
\eea
where  $\tau_2$ is a Pauli
matrix, so $i \tau_2$ provides the antisymmetric SU(2) contraction.

\begin{figure}[bht]
\unitlength.5mm
\begin{center}
\epsfig{file=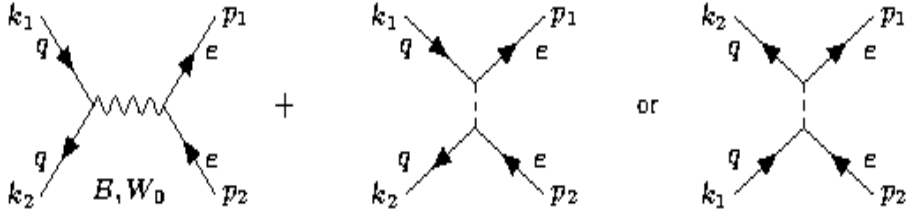,height=3.5cm,width=14cm}
\end{center}
\caption{\label{fig:Feynman}
Possible  SM and leptoquark  diagrams
for    $q \bar{q} \to e^+e^-$,
in the limit of neglecting $m_Z$.
  Only one leptoquark
diagram will be present at a time; the central
diagram is for fermion number $F= 0$ leptoquarks (the SU(2) doublets), the
last diagram for fermion number $F=2$ leptoquarks (the SU(2)
triplets and singlets). Momenta $k$ enter the graph, and
momenta $p$ leave.
 }
\end{figure}

These leptoquarks can contribute to $ q \overline{q} \to e^+ e^-$
(where $q \in \{u,d\}$) via $t$-channel exchange. The diagrams,
and interfering SM processes are given 
in figure \ref{fig:Feynman}.
It is convenient to write the matrix element ${\cal M}
(\overline{q} q_X\to e^+e_Y^-) $ 
as a spinor contraction $ {\cal T}_{XY}$ multiplied by a 
propagator $ {\cal P}_{q_X e_Y}$.
For instance, 
for $S_o$ with coupling $\lambda_R$,  the spinor
contraction after
a Fiertz transformation can be written
${\cal T}_{RR} =
  ( \overline{u} \gamma^{\mu}P_R u)
    (\overline{e} \gamma_{\mu} P_R e)$.
For $ XX \in \{ LL, RR \},$ or  $XY \in \{LR, RL\}$
\bea
\overline{|{\cal T}_{XX}|^2} = 
\hatu^2 ~~~,~~~
\overline{|{\cal T}_{XY}|^2} = 
\hatt^2  ~~~, ~~~~(\hatu = -2p_2\cdot k_1 ~,~ \hatt = -2p_1\cdot k_1)
\label{19ounon}
\eea
where the bar indicates an average over incident colour
and spin,  the momenta are as in figure
\ref{fig:Feynman}, and
\beq
\frac{d \sigmahat }{ d \hatt} =  
\frac{|\overline{{\cal M}}|^2}{16\pi \hat{s}^2 }
~~{\rm with} ~~\overline{|{\cal M}|^2} = \overline{|{\cal T}|^2} |{\cal P}|^2 
~~~.\label{defnP}
\eeq

The contact interaction analysis of
ATLAS is at $\hats >(400 $GeV)$^2$, so
we neglect $m_Z$, and  propagate the
massless $B$ and $W^0$.
Then the  propagators of  figure
\ref{fig:Feynman}  give
\beq
i {\cal P}_{q_X e_Y} = {\Big (}
Y_{e_Y} Y_{q_X } g^{'2} + T_{e_Y}T_{q_X } g^{2} 
{\Big ) }\frac{1}{\hats}
- (-1)^{F/2}\frac{\lambda^2}{2 (m_{LQ}^2-\hat{\tau} )} (\times 2)
\label{P}
\eeq
where $F$ is the leptoquark fermion number which
is 2 for doublet leptoquarks and zero otherwise,   
 $\hat{\tau}= \hatt$  for $F = 0$ leptoquarks
and  $ \hatu$ for $F = 2$,
and the $(\times 2)$ applies only  in the case of
triplet leptoquark exchange coupled to $d$ quarks.
Recall  that the  hypercharge and SU(2) quantum numbers 
of SM fermions are
$$
Y_{e_L} = -\frac{1}{2}~~,
Y_{e_R} = - 1~~,
Y_{q_L} = \frac{1}{6}~~,
Y_{u_R} = \frac{2}{3}~~,
Y_{d_R} = -\frac{1}{3}~~,
T_{e_L} =T_{d_L} =  -\frac{1}{2}~~
T_{u_L} = \frac{1}{2}~~.
$$
For each leptoquark, 
the  Fierz-rearranged  $\bar{q} q e^+e^-$vertices, with the
appropriate 
 propagators representing $Z/\g$ and
leptoquark exchange, are given in colomn two of table \ref{4fv}.
To obtain the contact interaction mediated by
a leptoquark, $\tau \to 0$
in this table and the SM part of the propagator should be dropped.
In most cases, the coefficient of the contact
interaction is $\lambda^2/(2m_{LQ}^2)$.

ATLAS bins its data in
  $\hats = M^2_{\epem}$ (see eqn \ref{M2ee}),
so it is  convenient to  express the total
 cross-section for  $pp \to  e^+e^-$  as
\bea
\sigma = 
2 \sum_{q=u,d}
\frac{1}{s} \int 
d \hats \, d\eta^+
\, d\hatt
 f_q(x_1)  f_{\bar{q}}(x_2) 
\left[
\frac{d \hat{\sigma}_{Z/\g}}{d\hatt}  
+ \frac{d \hat{\sigma}_{NP}}{d\hatt}  \right]
\label{sigint}
\eea
where $f_q$ is the  parton distribution function (pdf) for
the quark $q$ in the proton,
$x_1 = \frac{M_{\epem}}{\sqrt{s}} e^{\eta_+}$  
and $ x_2 = \frac{M_{\epem}}{\sqrt{s}} e^{-\eta_+}$ 
 are the fraction  of the proton's momentum carried 
by the parton,   there is a 2 because the
valence quark could be in either incident proton, 
  and the cross-section is separated into the
gauge boson mediated part plus  a New Physics part. 
The integration limits on 
$\eta_+ =(\eta_{e} + \eta_{\bar{e}})/2 $
and $\hatt$  should be determined from
the experimental cuts on the rapidities 
$\eta_{e^+},\eta_{e^-}$ of the $\epem$, however,
for simplicity, we integrate over the whole phase space.

\section{Leptoquark exchange  as a contact interaction}
\label{sec:CI}

In this section, we approximate leptoquark exchange as a contact interaction,
and try to constrain the leptoquark-mediated contact interactions
from the ATLAS analysis.  The challenge will be to deal with
the  different flavours and chiralities of the leptoquark-induced  operators,
which will affect the number of New Physics events, and the
distribution in $M^2_{e^+e^-}$.To see this,
the total cross-section of eqn (\ref{sigint}) can be written
\bea
\frac{d \sigma (pp\to e^+e^-)}{d \hats} 
&= & C(\hats) \frac{\hats}{s} \left( \frac{3g^4}{32 \hats^2}
+ \epsilon_{int}\frac{ g^2}{ \hats}
\frac{\lambda^2 }{2m_{LQ}^2 }
+ \epsilon_{NP}\frac{\lambda^4 }{4m_{LQ}^4} \right)~~~,
\label{use?}
\eea
where   the NP part is divided into
 interference-with-the-SM,  and with itself.
$C(\hats) $ is dimensionless, fixed by the electroweak
interactions of the quarks and electrons,  and includes
an integral over the  pdfs. The $\epsilon$s are
also dimensionless, and satisfy 
$-\sqrt{3/8} \leq \epsilon_{int}/\sqrt{\epsilon_{NP}}\leq \sqrt{3/8}$.
The  magnitude of $\epsilon_{NP}$  will depend on
which flavours of quark  couple to a given leptoquark,
and the magnitude and sign of $\epsilon_{int}$ will
depend on the flavour and chirality
of the participating fermions, 
because these control the interference  with the  $\gamma/Z$.

To analytically  compare   the $\epsilon$s
induced by leptoquarks,
to those arising from the ATLAS
contact interaction,  we suppose a simplistic
weighting of parton distribution functions (pdfs) in the proton, such that
\beq
\frac{f_u(x_1)f_{\bar{u}}(x_2) }{f_u(x_1)f_{\bar{u}}(x_2) + f_d(x_1)f_{\bar{d}}(x_2)} = \frac{2}{3} 
~~~,~~~
\frac{f_d(x_1)f_{\bar{d}}(x_2) }{f_u(x_1)f_{\bar{u}}(x_2) + f_d(x_1)f_{\bar{d}}(x_2)} = \frac{1}{3}~~~.
\label{poid}
\eeq
This approximation  will allow to estimate
 $\epsilon_{int}$ and $ \epsilon_{NP}$
from the partonic
matrix elements (given in the second colomn of table
\ref{4fv}).  

To make such estimates, we first  schematically write 
 the partonic 
 $Z/\g$-exchange cross section, multiplied by
pdfs,  as:
\bea
{\rm pdfs} \times
\frac{d\hat{\sigma}}{d\hatt}&=&
\frac{\hatt^2}{16\pi \hats^4}
\left[ g^4 T_e^2 (f_u f_{\bar{u}}T_u^2 + f_d f_{\bar{d}}T_d^2) +
2 g^2 g^{'2} T_eY_{e_L}[T_uY_{u_L}f_u f_{\bar{u}} +T_dY_{d_L}f_d f_{\bar{d}}]
\right.
\nonumber\\
&&~~~~~~~~~~~~~~~~~~~ +\left.
g^{'4}(Y_{e_L}^2 + Y_{e_R}^2)(f_u f_{\bar{u}}[Y_{u_L}^2 + Y_{u_R}^2] +f_d f_{\bar{d}}[ Y_{d_L}^2 + Y_{d_R}^2])
\right]~~~,
\nonumber
\eea
where inside the square brackets is the pdf-weighted
``propagator'' $|{\cal P}|^2$ of eqn (\ref{P}),
mutiplied by $\hats^2$.
With the approximations
 $s_W^2= 1/4$,  $g' = g/2$, and
eqn (\ref{poid}):
\bea
\frac{d {\sigma}_{Z/\g}}{d\hats} & \simeq&
\frac{3}{32}
\frac{g^4}{ \hats}
\frac{1}{48\pi s}
\int d\eta_+ (f_u f_{\bar{u}} + f_d f_{\bar{d}}  ) ~~~~ \Rightarrow  ~~~~~~
 C (\hats)  =  \frac{1}{48\pi }
\int d\eta_+ (f_u f_{\bar{u}} + f_d f_{\bar{d}}  )
\label{fondh}
\eea 
where the pdf integral over $\eta_+$ 
is a constant that we do not need to know.
(We also neglected experimental cuts in 
integrating over $\hatt$, which is hopefully
an acceptable approximation because the $\hatt$
dependence of $|{\cal T}|^2$  is common to the SM
and NP, and there is no $\hatt$ dependance of the
propagators in the contact approximation.)

 The ATLAS analysis \cite{ATLAS}  
follows  pythia\cite{pythia6} in  summing over  $u$ and $d$ flavours,
and restricts to doublet (``left-handed'') quarks. 
Using again the approximation (\ref{poid}),
and identifying $4\pi/\Lambda^2 =\lambda^2/2m^2$,
the  cross-section 
in the presence of the ATLAS contact interaction,
with constructive interference with the SM, 
can be approximated as
\bea
\frac{d {\sigma}_{CI}}{d\hats}
& \simeq&\frac{d {\sigma}_{Z/\g}}{d\hats} +
\left(\frac{1}{6}\frac{g^2}{\hats}
\frac{\lambda^2}{2 m^2} + \frac{\lambda^4 }{4m^4}\right)
\frac{\hats }{48 \pi s} \int d\eta_+ (f_u f_{\bar{u}} + f_d f_{\bar{d}}  )
\label{sighCI}
\eea
so $\epsilon_{int} =1/6$ and $\epsilon_{NP} = 1$ for the ATLAS
contact interaction.

\renewcommand{\arraystretch}{2}

\begin{table}[htb]
\hspace{-.5cm}
$
\begin{array}{||l|l|c|c|c|l||}
\hline
{\rm interaction}  & {\rm
 Fierz-transformed ~} {\cal M}   & \epsilon_{int} &
\epsilon_{NP}&
\lambda^2 \!
\left( \frac{(2{\rm TeV})^2}{m_{LQ}^2}\right) \! < & CL \\ \hline
ATLAS&
\begin{array}{c}
+,-\left[ (\overline{u} \gamma^{\mu} P_Lu)
(\overline{e} \gamma_{\mu} P_L e)
\left(\frac{|\lambda|^2}{2m^2}  - 
\frac{1}{4}
\frac{g^{2}}{\hats}  \right) \right.\\
~~~~\left. + (\overline{d} \gamma^{\mu} P_Ld)
(\overline{e} \gamma_{\mu} P_L e)
\left(\frac{|\lambda|^2}{2m^2}  +
\frac{1}{4}
\frac{g^{2}}{\hats} \right) \right]
\end{array}
&{ - \frac{1}{6} ,+ \frac{1}{6} }
 &{1}& 
1.1,0.73 
&
76\%,  87\%  
\\
\hline
%
%
( \lambda_{LS_o} \overline{q^c} i\sigma_2 \ell +
 \lambda_{RS_o} \overline{u^c} e) S_o^{\dagger} 
&
 ( \overline{u} \gamma^{\mu}P_R u)
    (\overline{e} \gamma_{\mu} P_R e)  
\left( \frac{|\lambda_{R}|^2}{2(m_o^2- \hat{\tau})}
 -\frac{2}{3} \frac{g^{'2}}{\hats}\right)
&
-\frac{2}{9}
& \frac{2}{3}
&.87& 89\%
\\
&
 (  \overline{u}  \gamma^{\mu}P_L u )(\overline{e} \gamma_{\mu}P_L e) 
\left(\frac{|\lambda_{L}|^2}{2(m_o^2- \hat{\tau})} 
- \frac{1}{4} \frac{g^{2}}{\hats}   \right)
& -  \frac{1}{3}
& \frac{2}{3}
&.58 & 98\%
\\  \hline
%
%
\lambda_{R\tilde{S_o}}  \overline{d^c} e \tilde{S}_o^{\dagger} 
& 
 (  \overline{d}\gamma^{\mu}  P_R d)(\overline{e} \gamma_{\mu} P_R e)  
\left(\frac{|\lambda_{R}|^2}
{2(\tilde{m}_o^2- \hat{\tau})} +
\frac{1}{3} \frac{g^{'2}}{\hats}  \right)
& \frac{1}{18}
& \frac{1}{3}
& 2.2 & 99\%
\\ \hline
%
%
(\lambda_{L} \overline{u} \ell + \lambda_{R}
\overline{q} i\sigma_2   e) S_{2}^{\dagger} 
&
 ( \overline{u} \gamma^{\mu} P_Ru)
(\overline{e} \gamma_{\mu}P_L  e) 
\left(
-\frac{|\lambda_{L}|^2}{2(m_{2}^2- \hat{\tau})}
-\frac{1}{3} \frac{g^{'2}}{\hats}\right) 
&\frac{1}{9} &
\frac{2}{3} 
&1.1 & 95\%
\\
& 
( \overline{u} \gamma^{\mu} P_L u)
(\overline{e} \gamma_{\mu} P_R e) 
\left(- \frac{|\lambda_{R }|^2}{2(m_{2}^2- \hat{\tau})}
-\frac{1}{6} \frac{g^{'2}}{\hats} \right) 
&
&
& &
\\
&
~~~
 + ( \overline{d} \gamma^{\mu} P_L d)
(\overline{e} \gamma_{\mu} P_R e) 
\left(- \frac{|\lambda_{R}|^2}{2(m_{2}^2- \hat{\tau})}
-\frac{1}{6} \frac{g^{'2}}{\hats} \right) 
&\frac{1}{12}
& 1 
&1.46 &  99.96\%
\\
 \hline
%
%
\lambda_{L \tilde{S}_{2}} \overline{d} \ell  \tilde{S}_{2}^{\dagger}
&
( \overline{d} \gamma^{\mu} P_R d)
 (\overline{e} \gamma_{\mu} P_L e)  
\left(-\frac{| \lambda_{L}|^2}{2(\tilde{m_{2}}^2- \hat{\tau})}
+ \frac{1}{6} \frac{g^{'2}}{\hats} \right) 
&-\frac{1}{36}
&\frac{1}{3}
& 6.6  (1.64) & 100\% ( 80\% )
\\  
\hline
%
%
\lambda_{LS_1}  \overline{q^c} i\sigma_2 \vec{\sigma} \ell 
\cdot \vec{S_1}^{\dagger}
&
(\overline{u} \gamma^{\mu} P_Lu) 
(\overline{e} \gamma_{\mu} P_L e)
\left(\frac{|\lambda_{L }|^2}{2(m_1^2- \hat{\tau})}
-\frac{1}{4} \frac{g^{2}}{\hats} \right) 
&&&&\\
&~~~~~~
+ (\overline{d} \gamma^{\mu} P_Ld)
(\overline{e} \gamma_{\mu} P_L e)
\left(\frac{|\lambda_{L }|^2}{(m_1^2- \hat{\tau})}
+  \frac{1}{4} \frac{g^{2}}{\hats} \right) 
&\simeq 0   & 4/3
&.80 & 86\%
\\  \hline
\end{array} $
\caption{ Fierz-transformed two-electron-two quark  matrix
elements  induced by the leptoquark, $\gamma$ and
$Z$ exchange  diagrams of figure \ref{fig:Feynman},
in the limit $m_Z \to 0$. (possible quark-neutrino
interactions are not included). 
 $ \hat{\tau}$ can be $\hatt$ or $\hatu$.
 The third and fourth   colomns 
 estimate  the coefficients in
eqn (\ref{use?}), using the approximation
of eqn (\ref{poid}). The second last colomn is the
bound on $\lambda^2$, for $m_{LQ}= 2$ TeV,
 assuming  the ATLAS
limits on contact interactions can be
translated to leptoquarks using eqn (\ref{conserv}).
The last colomn is an estimate of the confidence
level (see eq. \ref{CL}) of that bound, obtained with the cross-section
of eqn (\ref{use?}).
\label{4fv}}
\end{table}

\renewcommand{\arraystretch}{1}

\begin{figure}[hbt]
\unitlength.5mm
\begin{center}
\epsfig{file=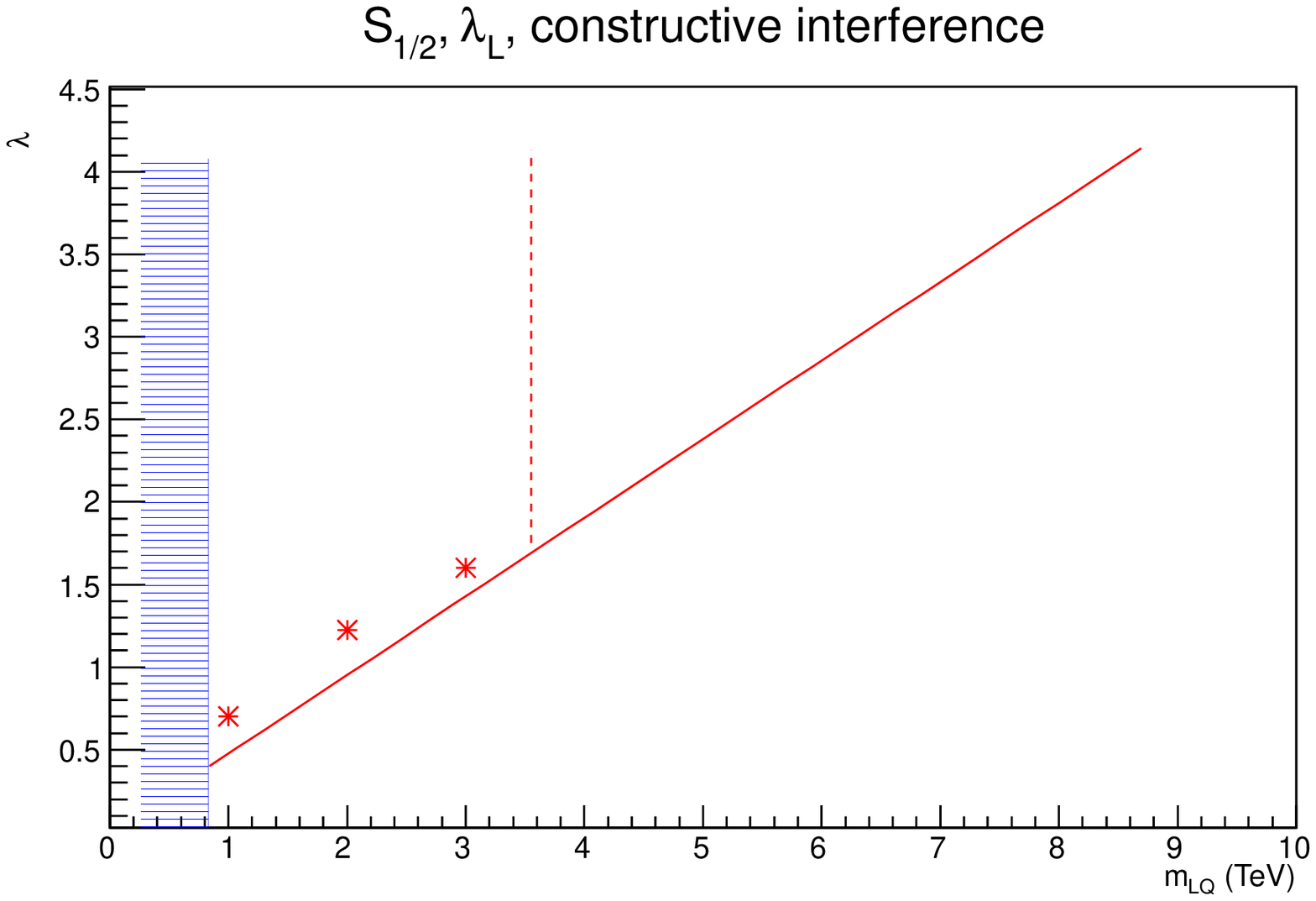,height=5cm,width=8cm}
\epsfig{file=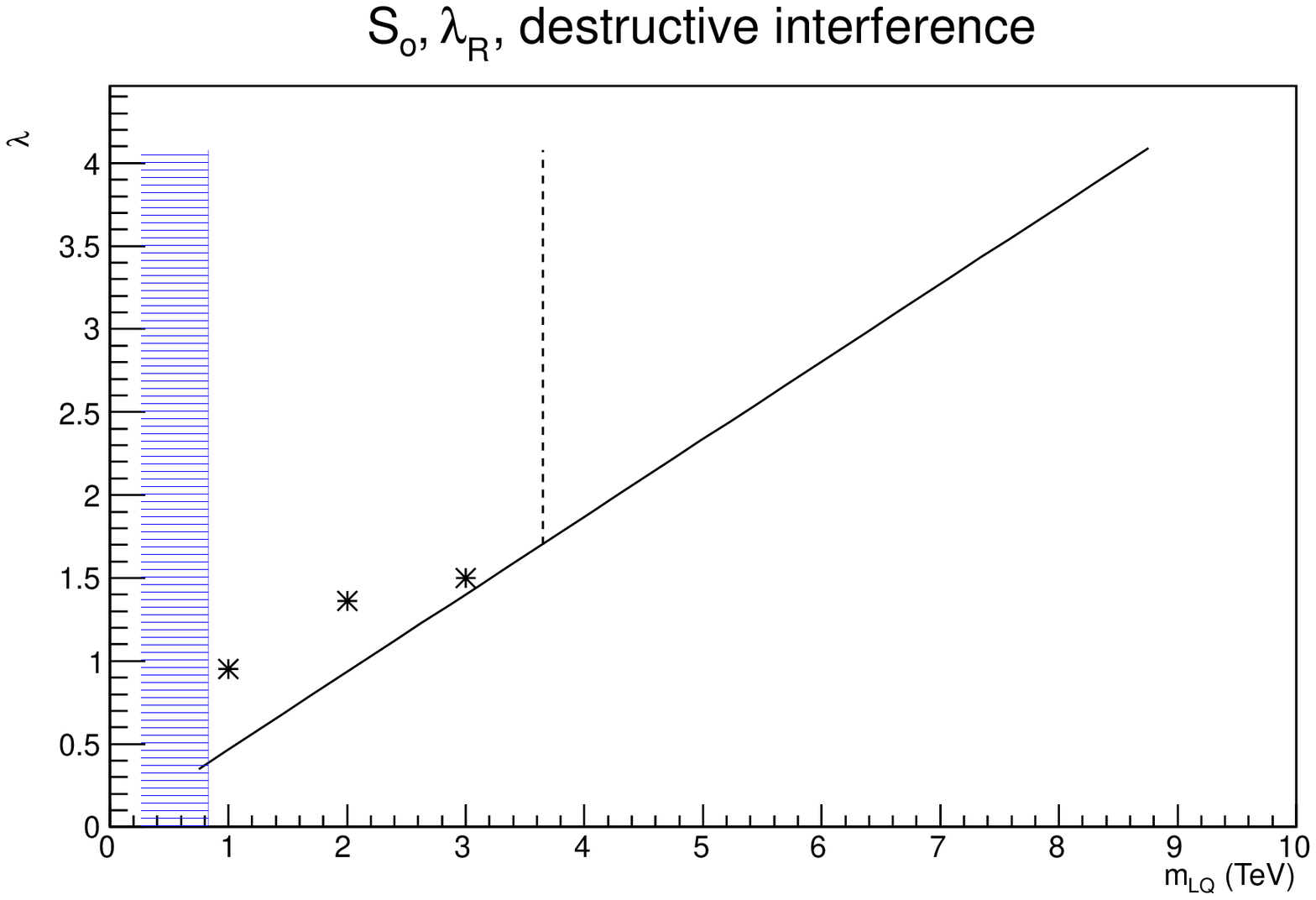,height=5cm,width=8cm}
\end{center}
\caption{ The parameter space above the diagonal line
is excluded, if leptoquark exchange can be
described as a contact interaction (see table
\ref{4fv}). Left plot
for 
$S_{2}$ with coupling $\lambda_L$, right
plot for $S_o$ with  coupling $\lambda_R$.
 The blue region
to the left  with horizontal hashes is
excluded by  CMS searches\cite{CMS} for  pairs
of first generation leptoquarks.
 The contact interaction approximation 
should  apply
to the right of the dashed line;
if the leptoquark propagator
is taken into account, only the
region above the stars is excluded (see section
\ref{sec:prop}).
\label{fig:tri} }
\end{figure}

The contact interactions induced by the
various leptoquarks  differ from the one studied
by ATLAS, as
can be seen from  the second colomn of table \ref{4fv}.
The values of $\epsilon_{int}$ and $\epsilon_{NP}$
can be estimated, as above,  and are  given  in 
 table  \ref{4fv}. One sees that the  
leptoquarks with constructive interference (positive
$\epsilon_{int}$) have a smaller
$\epsilon_{int}/\sqrt{\epsilon_{NP}}$ ratio  
than the  ATLAS  operator. So one could hope to
constrain these leptoquarks by simply rescaling the
ATLAS bound.  We conservatively
 rescale the bound as
\beq
\left.
\epsilon_{int} \frac{4\pi}{\Lambda^2} \right|_{ATLAS} \geq  
\epsilon^{LQ}_{int} \frac{ \lambda^2}{2 m_{LQ}^2}
~~~~~~~~
\label{conserv}
\eeq
where $ \Lambda_{cons} = 11.7$  TeV and
$\epsilon_{int} = 1/6$ on
 the left side. 
For the leptoquark $S_{2}$ with coupling $\lambda_L$,
this excludes above the  red  diagonal line 
of the left figure \ref{fig:tri}. 

Analytic estimates suggest that  eqn (\ref{conserv}) is conservative:
if the bound arises from the  interference term, then one expects
$\epsilon_{int}^{ATLAS} \frac{4\pi}{\Lambda^2} \simeq 
\epsilon_{int}^{LQ}   \frac{ \lambda^2}{2 m_{LQ}^2}$.
However, if the bound
came from the $|NP|^2$ term, then
one might expect $\sqrt{ \epsilon_{NP}^{ATLAS}}
\frac{4\pi}{\Lambda^2} \simeq   \sqrt{\epsilon_{NP}^{LQ}} \frac{ \lambda^2}{2 m_{LQ}^2}$, which would give a stronger
bound on the leptoquark couplings.  For  the ATLAS  contact interaction,
and leptoquark-induced contact interactions with
$\epsilon_{int}/\sqrt{\epsilon_{NP}} \simeq 1/6$,
the $NP^2$ term becomes larger than the
interferences at $\sqrt{\hats} \gsim 900$ GeV,
so  dominates 
 the last bin with data (= 1200 TeV $<\sqrt{\hats} <1800$ TeV).

For leptoquarks that have  destructive  interference
with the SM, the ratio
$\epsilon_{int}/\sqrt{\epsilon_{NP}}$ departs significantly 
from the ratio of the ATLAS operator. Nonetheless
  we again  take the translation
rule of eqn (\ref{conserv})
with $\Lambda_{des} \geq 9.3 $  TeV.
On  the right in  figure \ref{fig:tri}
is plotted the exclusion for  $S_o$ with coupling 
$\lambda_R$.
In the second last colomn of figure \ref{4fv},  eq.
(\ref{conserv}) is used to translate the ATLAS
bound to all the singlet and doublet leptoquarks.
The next subsection contains some  simple statistics
to support  eqn (\ref{conserv}).

A bound was estimated for  the triplet leptoquark $S_1$, 
for which the interference
almost vanishes, as
\beq
\left.
 \frac{4\pi}{\Lambda^2} \right|_{ATLAS} \geq  
\sqrt{\epsilon_{NP}} \frac{ \lambda^2}{2 m_{LQ}^2}
~~~~~~~~
\label{guess2}
\eeq
where $ 1/\Lambda^2 $ on the left side is
the average of the ATLAS constructive
and destructive bounds $ \frac{1}{2}
(\frac{1}{11.7^2} + \frac{1}{9.3^2})$,   and
$\epsilon_{NP} = 4/3$ on
 the right side. 
For the doublet $\widetilde{S}_{2}$, whose interference
with the SM is very suppressed,  the estimated bound
from eqn (\ref{conserv}) is to weak  to be interesting.
If instead, the
interference term is neglected, a bound of
$\lambda^2 \geq 1.64$ for $m_{LQ}= 2$ TeV can
be estimated from eqn (\ref{guess2}), 
(this is given in parentheses in the table).

\subsection{ Comparing partonic cross-sections to ATLAS data}

None of the leptoquarks induce the contact interaction
constrained by ATLAS, so it is not clear how to translate
the  ATLAS bound to leptoquarks. In  particular, the
shape in $\hats$  of the differential cross-section, eqn  (\ref{use?}),
 will depend on the different values of the $\epsilon$s. This will
change the number of New Physics events in the
ATLAS bins, and affect the  overall deviation from the SM.
 The aim of this subsection is to confirm that eqn (\ref{conserv}) is
conservative, using simple statistics and partonic
cross-sections.

We focus on the last three ATLAS bins in $\hats$
(see table \ref{tab:atlas}), and suppose that the
pdfs are decreasing fast enough  with increasing
$\hats$, that in each bin $b$,
 the number of signal  NP events  plus background
SM events  can be estimated as
\beq
  (r_b^{NP} -1) n(Z/\g) + n(SM) 
~~~,~~~r^{NP}_b \equiv  \frac{\sigmahat_{Z/\g+NP} (\hats)}
{\sigmahat_{Z/\g} (\hats)}
\eeq
where 
$r_b^{NP}$ is the ratio  of  
 the    $Z/\g+NP$ cross-section of eqn (\ref{use?})
to the   $Z/\g$ cross-section (this neglects experimental acceptances),
taken  at the left side of each
$\hats$  bin,  calculated for
$m_{LQ}=2$ TeV with the value of $\lambda^2$ give
in the second last colomn of table \ref{4fv},
and using the  correct  $\epsilon_{int}$ and $\epsilon_{NP}$
for each leptoquark.  
The 
$n(Z/\g)$ and  $ n(SM) $ are respectively
the number of $Z/\g$  events
and total number of SM events, expected by ATLAS
(see table \ref{tab:atlas}).

Then we estimate a ``confidance level'' $CL$ for the 
exclusion as follows. Consider first the case
of constructive interference, where the SM+NP cross-section
is larger than  the SM  cross-section. 
Then assuming Poisson statistics for 
the last three bins,  the probability of counting
less than or equal  the observed number  $N$ events, 
when expecting
$\nu$, is
\bea
P_b  &=&  e^{-\nu} \sum_{n \leq N}\frac{\nu^{n}} {n!}  ~~~.\nonumber 
\eea
The ``confidence level''  then is estimated as
\bea
CL = 1 -\frac{\Pi_b P_b (\nu = SM+ NP)}{\Pi_b P_b (\nu = SM)}
\label{CL}
\eea
In the case of  a leptoquark-mediated contact
interaction with destructive interference, the expected number of
SM+NP events in some bins is less than the expectation for
the SM alone. For those bins, $P_b$ is taken as the
probability of observing more than or equal 
to the observed number $N$.  These ``confidence levels''
are listed  in the last colomn of  table \ref{4fv},
for the bounds  quoted in the second last colomn
 (obtained from eqn (\ref{conserv}) and
(\ref{guess2})).
Our CL estimates are higher for the leptoquark limits
than for the ATLAS contact interaction, which reassures us
that our bounds are conservative. However, the variation
in the CLs indicates that eqn (\ref{conserv}) is not
a reliable approximation. To obtain a consistent
confidance level for various values of 
$\epsilon_{int}/\sqrt{\epsilon_{NP}}$ would require a more
sophisticated study.


\section{Including the leptoquark propagator}
\label{sec:prop}

The aim  of this section is to estimate the
consequences of including
 the leptoquark   propagator $1/(m^2_{LQ} - \hat{\tau})$
 (where $\hat{\tau} \in \{ \hatt,\hatu\}$, and
recall $\hatt, \hatu < 0$),
which only 
reduces to a contact interaction  in the 
$|\hat{\tau}| \lsim \hats \ll m^2_{LQ}$ limit. It is 
interesting to explore the    $ m^2_{LQ} \lsim \hats$
range, because  the lower bound
on the mass of pair-produced first generation leptoquarks 
is  830 GeV\cite{CMS}, whereas the highest  bin in the ATLAS
analysis is $\sqrt{\hats} > 1800$ GeV.

\begin{figure}[ht]
\unitlength.5mm
\begin{center}
\epsfig{file=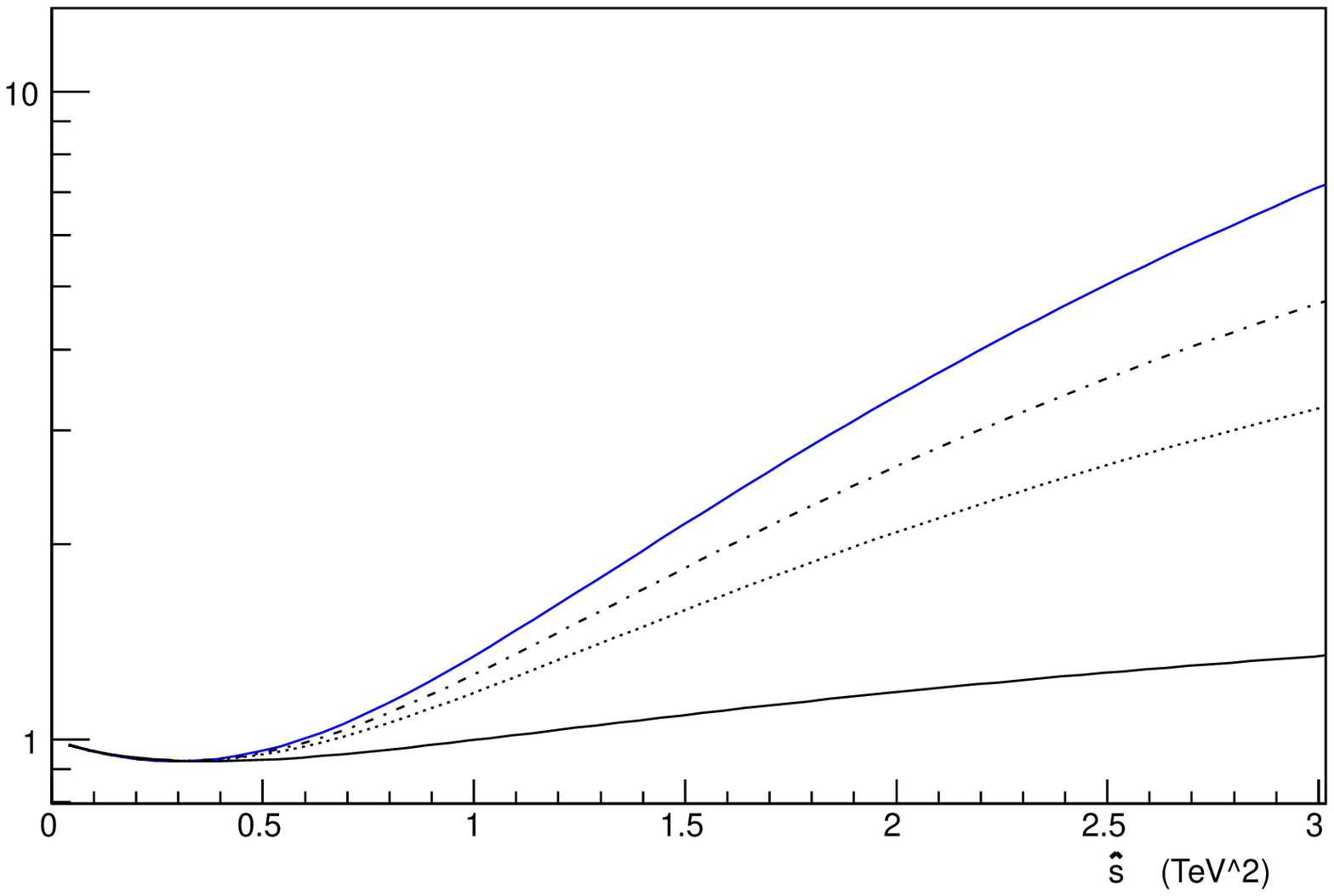,height=6cm,width=8cm}
\epsfig{file=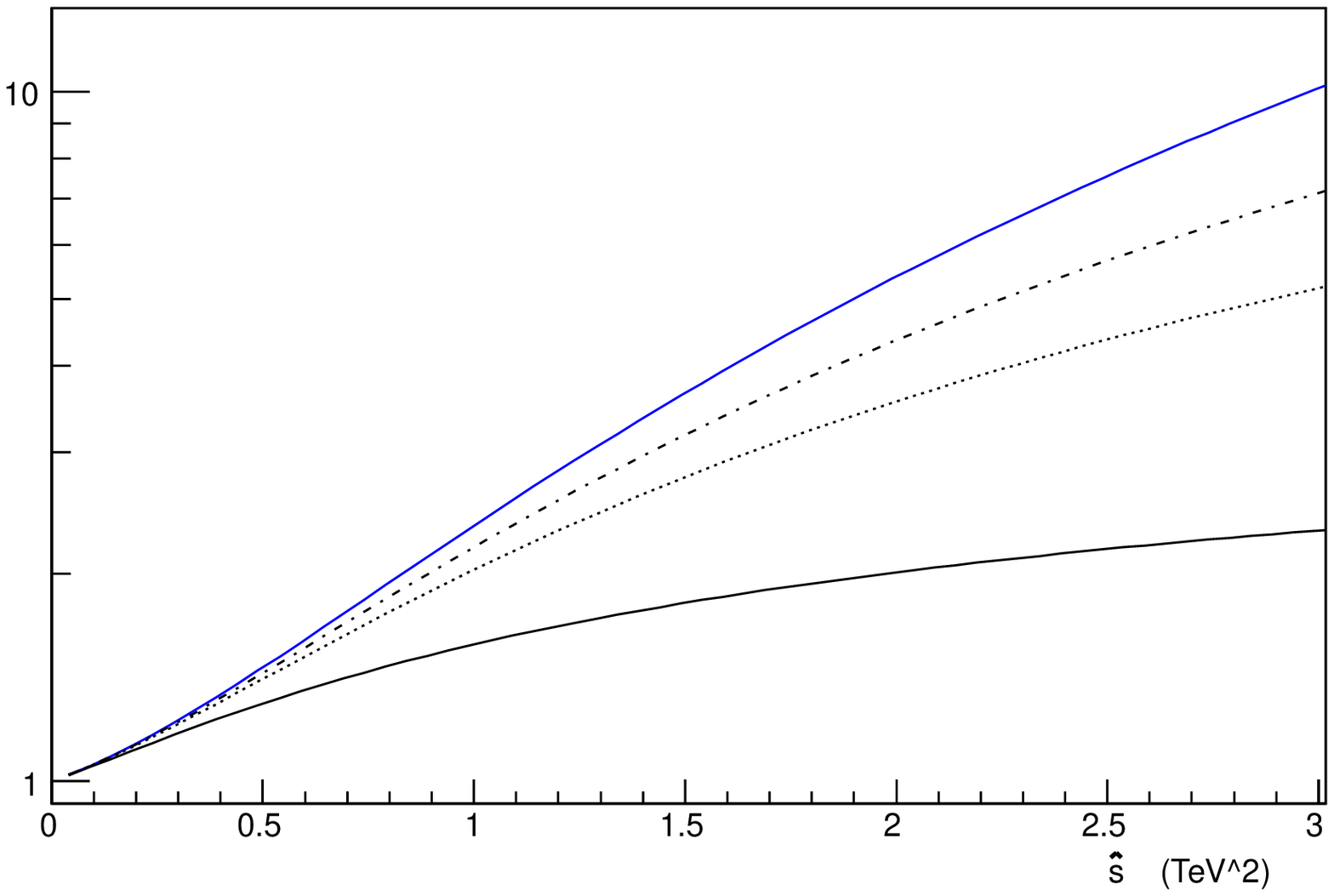,height=6cm,width=8cm}
\end{center}
\caption{ Ratios of $d\sigma/d\hats$(New Physics)
to $d\sigma/d\hats$($Z/\g$) --- see
 equations
(\ref{fondh}), (\ref{sighCI}) and  (\ref{sighLQ}). From
the lowest curve upwards, the New Physics
is  $Z/\gamma$  exchange plus 
a $t$-channel leptoquark   with  $m_{LQ}=$
1 (solid),2 (dotted) and 3 (dash-dotted) TeV,
and finally  $Z/\gamma$ 
exchange plus  a  contact
interaction
 with coefficient $\lambda^2/(2m^2)= 1/8$ TeV$^{-2}$
 (blue). The leptoquark
couplings are chosen to reproduce the
contact interaction in the $\hat{s}\to 0$ limit.  
On the left, destructive interference,
constructive on the right.
\label{fig:seceff} }
\end{figure}

The effect of the  massive LQ propagator 
 can be seen by comparing the
partonic cross-sections  for  LQ exchange
versus a contact interaction. With the approximation of
eqn(\ref{poid}), $d\sigma/d \hats$ for
the ATLAS contact interaction is given in eqn
(\ref{sighCI}). In the same approximation, with the
same values of $\epsilon_{int} = 1/6$ and
$\epsilon_{NP}=1$,  $d\sigma/d \hats$
for leptoquark exchange is
\bea
\! \! \! \! \! \! \!
\frac{d \sigma_{LQ}}{d\hats} &=& 
\frac{ d\sigma_{Z/\g}}{d\hats}+
\frac{3 C }{ s \hats}
\left[
\frac{g^2\lambda^2}{12}
\left(\frac{1}{2} -\frac{m^2}{ \hats} 
  +\frac{m^4}{\hats^2} \ln (1+\frac{\hats}{m^2}) \right)  +
\frac{\lambda^4}{4}
\left(\! \!1 -2\frac{m^2}{\hats} \ln (1+\frac{\hats}{m^2})   
+\frac{m^2}{(m^2 + \hats)}  \right)
\right]
\label{sighLQ}
\eea
In  figure \ref{fig:seceff} are plotted
the differential cross-sections for 
leptoquarks of masses 1, 2 and 3 TeV,
and the contact interaction they induce
(the leptoquark couplings are adjusted
such that all three give the same contact
interaction).
It is clear that  for
$ m_{LQ}^2> 4 \hats$,  the contact interaction
approximation  reproduces leptoquark exchange to within 20\%.
This is represented  in figure \ref{fig:tri} as
the  dotted vertical  lines,
to the right of which the contact interaction 
approximation is justified.

To constrain a leptoquark of $m_{LQ}\lsim 3.6 $ TeV
using the ATLAS  contact interaction analysis, we should  account
for the differences in  cross-section shape, as a function
of $\hats$.  For the leptoquarks   represented
in figure \ref{fig:tri}, we estimate the value
of $\lambda$ which can be excluded 
for masses of 1,2 and 3 TeV,  by requiring
that the estimated confidence level of the leptoquark
exclusion, exceed  the  CL for the ATLAS  contact interaction.
For instance, for $S_{2}$ with coupling $\lambda_L$, 
the  bounds in the contact interaction (CI) approximation,
and including the propagator are
\beq
{\rm CI:} ~~\lambda^2_L < \left\{
\begin{array}{cc}
.28 ~~, &m_{LQ} = 1~{\rm TeV}\\
1.1 ~~, &m_{LQ} = 2~{\rm TeV}\\
2.5 ~~, &m_{LQ} = 3~{\rm TeV}
\end{array}
\right.
~~,~~{\rm propagator:}~~
\lambda^2_L < \left\{
\begin{array}{cc}
.50 ~~, &m_{LQ} = 1~{\rm TeV}\\
 1.5 ~~, &m_{LQ} = 2~{\rm TeV}\\
 2.6  ~~,&m_{LQ} = 3~{\rm TeV}
\end{array}
\right.
\label{bds}
\eeq
The bounds  obtained with
the propagator are plotted as  stars in figure \ref{fig:tri}.
As expected, the effect of the leptoquark
mass, for $m_{LQ}^2 \lsim \hats$ is to
weaken the bound.

\section{Discussion}
\label{sec:disc}

A signal for a contact interaction of coefficient
$\lambda^2/m^2$  is a plateau at the
high energy end of a decreasing distribution, possibly
preceded by a valley in the the case of
destructive interference with the SM. For
$q \bar{q} \to \epem$, the exchange  of   $Z/\g$
is the  principle SM contribution,  responsable
for the cross-section decreasing as $1/\hats$.  So
one expects  sensitivity to
\beq
\frac{\lambda^2}{m_{LQ}^2} \gsim \frac{g^2}{M^2_{\epem , max}}
\eeq
where $M^2_{\epem , max}$ is the  $\epem$ invariant mass-squared
of the highest bin.  
The contact interaction approximation is expected to be
valid for
\beq
 \lambda^2 < 4\pi ~~~~,~~~~\hats \ll m_{LQ}^2 
\eeq
where the strong-coupling upper limit on $\lambda$
is approximately a unitarity bound,
and where we impose the  
$\hats \ll  m_{LQ}^2 $ condition as $ \hats \leq  m_{LQ}^2/4$
(from fig \ref{fig:seceff}).   So
contact interaction searches at colliders are sensitive to
a triangular area in $\lambda, m_{LQ} $ parameter space 
(above the diagonal line of figure \ref{fig:tri}). 
This illustrates
that  the ``contact interaction approximation'', where
the two parameters $\{\lambda, m_{LQ} \}$, are replaced by a
single parameter $\lambda^2/ m_{LQ}^2$, is not well-satisfied
at colliders.

In addition, there is another effect  to parametrise.
Many of the channels in which  contact interactions are
searched for  (for instance
 $qq \to qq$,  or $q\bar{q} \to \epem$ as studied here)
 can be mediated by the SM. In the  present case, 
both the SM and leptoquark cross-sections have
the same angular dependence (because
the four-fermion interactions are of
the form $(\overline{q}\g^\mu P_Xq ) (\overline{e}\g^\mu P_Y e)$) 
so the cross-section for leptoquark plus
$Z/\g$ exchange is of the form given in eqn (\ref{use?}):
$$
\frac{d\sigma}{d\hats}  \sim c(\hats) \left( \frac{3g^4}{32 \hats}
+ \epsilon_{int}\frac{g^2\lambda^2}{m^2}
+  \epsilon_{NP} \frac{\lambda^4 }{m^4}\hats
\right)
$$
where  the sign and magnitude of the interference
term (encoded in $\epsilon_{int}$)  depend on the flavours and chiralities of
the contact interaction.  If the interference
term is large and negative,   there will be a valley before the
plateau, if it is positive, there  will be excess 
events before reaching the plateau... and if it is
negative and small, the plateau can merely be delayed.
In general,  experimental bounds  are quoted on
a few  discrete choices of the ratio  $\epsilon_{int}/\epsilon_{NP}$.

The aim of this paper was twofold. First, to extract
bounds on leptoquarks from the LHC searches for
contact interactions in the process $pp\to \epem +X$.
Leptoquarks interacting with electrons and  first generation quarks 
could contribute to $q\bar{q} \to \epem$ via
$t$-channel exchange. In table \ref{4fv} are quoted
the resulting limits, obtained via some hopefully conservative
analytic arguments. The limits have varying confidance
levels, indicating the difficulty of translating
current bounds to leptoquarks. The limits in table \ref{4fv}
neglect the first effect discussed above, of the
the leptoquark propagator. For small leptoquark
masses ($\sim$ TeV), the effect of the propagator
is to weaken the bound on $\lambda^2$ by less
than a factor 2 (see eqn (\ref{bds}), and figure \ref{fig:seceff}).

The second aim of this paper was to explore whether
experimental bounds on a selection of contact interactions
can be readily translated to  New Physics scenarios. The answer is no. 
There are two issues which arise:
\ben
\item the flavour and chirality of the experimentally
bounded interactions is unlikely to correspond to 
most models (as in the case here, none of the leptoquarks
induce the contact interaction studied by ATLAS). 
A conservative and fastidious solution 
 would be  for the experimental collaborations
to set bounds on a complete set of contact interactions,
which do not interfere among themselves, but do induce
every possible  helicity amplitude.
An alternative (which we will explore in a subsequent
publication),  would be for the
experimental collaborations to set bounds on
$\epsilon_{NP}$ and $\epsilon_{int}$, that is,
perform a two parameter fit to 
the  coefficients of the interference
and $|NP|^2$ terms in the (partonic) cross-section. 
 Then for a given model,
one merely needs to estimate these
parameters, which can be done from partonic
cross-sections with naive approximations to the pdfs.  

Notice that table \ref{4fv} and eqn (\ref{bds})
suggest that this effect is more important
than the following one:  bounds vary more with changes
in the coefficient of the interference term, 
than when the propagator in taken into account.

\item The four-momentum of
the new particle mediating the ``contact interaction''
can only be neglected  if the new  particle is
very heavy and strongly coupled.
If the propagator $\sim 1/(\hatt - m^2)$ of a new particle exchanged
in the $t$-channel is retained, the contact interaction is suppressed,
because $\hatt \sim -\hats$.
In the case of leptoquarks, 
we estimated the this weakens the bounds  on $\lambda^2$
by at most 50\% (see eqn (\ref{bds})).
\een

In summary, we estimated that contact interaction
searches in  $pp \to \epem$ at the LHC can
exclude first generation leptoquarks  with couplings
$\lambda^2 \gsim m_{LQ}^2/(2 ~{\rm TeV})^2$ (see
table \ref{4fv} for specific bounds). Two difficulties
arose in translating the experimental bounds to
 all the possible leptoquarks. The most significant problem
is that the sign and size of the interference
with the SM varies from one leptoquark to another,
and significantly affects the shape of cross-section and
therefore the bounds. Secondly, the
leptoquarks are rarely heavy enough to
justify neglecting their four-momentum in
the propagator, which, when included,
can  weaken the bound on $\lambda^2$  by < 50\%. To address
the first problem,  it could be interesting if the experimental
collaborations set simultaneous bounds on
the contact interaction coefficient $4\pi/\Lambda^2$,
and on the size and magnitude of the interference
with the SM.


\end{document}